\documentclass{article}
\usepackage{spconf,amsmath,amssymb,graphicx,booktabs,url}
\usepackage[hidelinks]{hyperref}

\newcommand{\WAR}{\ensuremath{\mathrm{WAR}}}

\title{Oracle Complementarity Is Not Realizable Complementarity\\
in Frozen-Encoder Audio--Visual Emotion Recognition}

\name{Benjamin Hurt}
\address{University of California, Santa Barbara}

\begin{document}
\maketitle

\begin{abstract}
Complementarity analyses of multimodal systems commonly report an
\emph{oracle} ceiling (the fraction of examples on which at least one
unimodal branch is correct) and interpret the gap between it and realized
fusion accuracy as recoverable headroom. We show that this ceiling is not
a fusion target. Using frozen self-supervised audio and visual encoders
with a trained head, we measure the ceiling and realized gain on CREMA-D
under two audio encoders (one whose fine-tuning lineage includes CREMA-D,
one clean) and on EAV under the contaminated encoder. Replacing the
fine-tuned encoder with a self-supervised one triples the headroom, from
$+0.055$ to $+0.186$. A second clean encoder and a controlled degradation
of the contaminated one fall on the same curve, so provenance moves the
headroom by changing audio-branch accuracy. Concatenation realizes at most about
half of the headroom (converted fraction $0.55$), and on EAV the converted
fraction is indistinguishable from zero despite larger headroom. A learned
router is beaten by plain concatenation everywhere. Oracle headroom measures branch disagreement,
not a fusion budget.
\end{abstract}

\begin{keywords}
multimodal fusion, emotion recognition, self-supervised representations,
complementarity, dataset contamination
\end{keywords}

\section{Introduction}
\label{sec:intro}

Frozen self-supervised encoders make multimodal affect systems cheap and
lightweight to assemble: a pretrained speech encoder, a pretrained face
encoder, and a small trained head reach competitive accuracy without
fine-tuning either backbone. The common rationale for investing in the
fusion mechanism itself is the \emph{oracle} ceiling, the accuracy of a
hypothetical selector that always picks a correct branch when one exists.
Oracle--fusion gaps of five to fifteen points of accuracy are commonly
read as headroom that could be recovered through a better gate, router,
or attention mechanism. A related inclination drives recent in-domain
work on modality scheduling, which dynamically prioritizes
under-contributing modalities to better realize their
contribution~\cite{modeq2025}.

The oracle is attractive because it is cheap. It requires no training,
only two sets of predictions. Although alternatives exist (taxonomies of
fusion strategy~\cite{baltrusaitis2019multimodal}, information-theoretic
decompositions of redundancy, uniqueness, and
synergy~\cite{liang2023quantifying}), the oracle remains a motivating
statistic in applied multimodal papers precisely because it is a single
number with an operational reading. We argue that this operational
reading is unsound, and that the error is structural, not a matter of
insufficiently clever fusion. The oracle counts clips on
which the branches \emph{disagree in the right direction}. A deployable
system must additionally \emph{identify and act on} those clips from the
inputs alone. A complementarity analysis that reports large headroom therefore
invites exactly the fusion machinery (routers, gates, cross-attention)
our results show does not close it.

The gap between the oracle and realized selection is long documented in
dynamic classifier selection, where the oracle is the standard upper bound
and selectors often fail to pick a competent classifier even when one
exists~\cite{cruz2018dcs,souza2017oracle}. Concurrent work outside speech
reaches related conclusions by other routes. In language-model routing, the recorded oracle has been
decomposed into a reproducible ceiling and a non-recoverable component
attributed to single-draw decoding noise~\cite{routinggap2026}, and
co-failure analyses across pools of models bound what any router can
capture~\cite{cofailure2026}; in multimodal retrieval, per-sample
modality relevance has been separated from modality
utility~\cite{modutility2026}. Our double-fault floor (Sec.~\ref{sec:defs})
is the same co-failure ceiling concept applied to modality branches
rather than to pools of language models. Our contribution is not to introduce it but to show
what it implies here. With deterministic predictions and fixed labels,
neither the decoding-noise nor the query-relevance mechanism applies, and
the unrealized headroom is disagreement that none of the mechanisms we
test converts into accuracy at inference.

We measure oracle headroom against realized fusion gain on CREMA-D under
a contaminated and a clean frozen audio encoder, and on EAV under the
contaminated one.
Replacing the
heavily fine-tuned encoder with a self-supervised one multiplies the
headroom by $3.4\times$, and a controlled degradation of the contaminated
encoder shows why. Headroom follows audio-branch accuracy, so
contamination hides it by inflating the stronger branch. The fraction that
concatenation converts stays near half at best. It rises from $0.39$ to $0.55$ on
CREMA-D, and on EAV, whose headroom is $2.5\times$ CREMA-D's contaminated
one, it is indistinguishable from zero. A learned router
underperforms plain concatenation in every corpus and regime tested,
including the clean regime where headroom is largest. The contribution is
the diagnosis, not a fusion method. Oracle headroom should be read as a
property of branch accuracy and disagreement, not a recoverable
budget.

\section{Experimental Setup}
\label{sec:setup}

\noindent\textbf{Corpora.}
CREMA-D~\cite{cao2014cremad} contains 7442 acted studio clips from 91
actors over six emotion categories.
EAV~\cite{lee2024eav} contains 4200 clips from 42 subjects over five
categories and is elicited in a more naturalistic conversational
protocol.

\noindent\textbf{Encoders.}
All encoders are frozen feature extractors; only a small MLP head is
trained. The audio branch is emotion2vec+ base or the self-supervised
non-plus emotion2vec, both 768-d utterance
embeddings~\cite{ma2024emotion2vec,ma2024emotion2vecplus}.
As a provenance robustness check (Sec.~\ref{sec:results}) we add
XEUS~\cite{chen2024xeus}, a clean, architecturally distinct 1024-d encoder.
The visual branch is EmoNet (256-d)~\cite{toisoul2021emonet}, MAE-DFER
(512-d)~\cite{sun2023maedfer}, or their concatenation, which we call the
visual ensemble and use for all headline results.
Published CREMA-D results for these encoders are obtained by fine-tuning
them ($77.4$ WAR video-only for MAE-DFER, $84.9$ for the audio-visual
HiCMAE~\cite{sun2024hicmae}), and no frozen-encoder audio-visual
baseline exists for this corpus, so our absolute accuracies are not
comparable to those figures.
Fusion is at the embedding rather than the posterior level, chosen by a
preliminary pooling and normalization sweep.

\noindent\textbf{Protocol.}
We follow the frozen-encoder, lightweight-head convention established for
self-supervised speech representations~\cite{yang2021superb}.
Evaluation is subject-independent GroupKFold over actor identity, with five folds on CREMA-D and seven on EAV and zero train/test actor overlap
verified per fold.
Every reported quantity is computed per seed over all out-of-fold clips
and then averaged over seeds $\{0,1,2\}$, and every clip count is a mean
over the three seeds. Confidence intervals are 95\% actor-level cluster
bootstrap intervals ($10{,}000$ resamples); quantities computed from a single seed's
out-of-fold posteriors are marked as such.
Head training uses deterministic CUDA reductions; Table~\ref{tab:main},
the routing comparisons, and the degradation sweep are verified
reproducible across processes.
Accuracy is weighted average recall (WAR).

\noindent\textbf{Contamination and the design grid.}
Benchmark contamination, evaluation data appearing in a model's
training or fine-tuning corpus, is a recognized threat to reported
gains~\cite{sainz2023contamination}. The frozen emotion2vec+ base encoder is fine-tuned on a large
undisclosed pseudo-labeled pool, initialized from a seed model that was
itself fine-tuned on the EmoBox collection~\cite{ma2024emobox}, whose 32
corpora across 14 languages include CREMA-D. CREMA-D can thus enter the
audio branch through this two-hop pathway, so the CREMA-D audio branch is
optimistic and subject-independent cross-validation at the head level
does not remove encoder-level leakage.
EAV is not an EmoBox member, and the visual encoders are unaffected in both cases. The visual ensemble scores $0.652$ identically under both
audio encoders, which we use as a control.
We report CREMA-D under both the contaminated (plus) and clean
(non-plus) encoders because published complementarity numbers resemble
the contaminated regime.
The clean encoder cannot be evaluated on EAV. EAV's Speaking condition
draws on only about 19 distinct scripted (trial\,$\times$\,emotion)
sentence templates, so a content-sensitive encoder decodes the script
rather than the prosody. Non-plus emotion2vec reaches $0.998$
subject-independent accuracy on EAV audio, against a text-only TF-IDF
ceiling of $1.000$ and a chance rate of $0.200$.
The design therefore has three cells, CREMA-D $\times$ \{contaminated, clean\} plus EAV $\times$ \{contaminated\}.

\section{Oracle, Realized Gain, and Unrealized Gap}
\label{sec:defs}

Let $a_i, v_i, f_i \in \{0,1\}$ indicate whether the audio branch, the
visual branch, and the fusion system respectively classify clip $i$
correctly, over $N$ clips. We define
\begin{align}
O   &= \tfrac{1}{N}\textstyle\sum_i \max(a_i, v_i), &
D   &= \tfrac{1}{N}\textstyle\sum_i (1-a_i)(1-v_i), \\
H   &= O - \WAR(a), &
\Delta &= \WAR(f) - \WAR(a), \\
U   &= O - \WAR(f), &
\rho &= \Delta / H .
\end{align}
$O$ is the \emph{oracle}, the accuracy of a selector with perfect
knowledge of which branch is correct, and hence an upper bound on any
late-fusion selector. $H$ is the oracle \emph{headroom} above the
stronger branch, $\Delta$ the \emph{realized gain}, $U$ the
\emph{unrealized gap}, and $\rho$ the \emph{converted fraction}. $D$, the
double-fault rate, is the floor on which no branch selector can
help~\cite{kuncheva2003diversity}; for two branches, $D = 1 - O$.
Embedding-level concatenation is not a selector and is not bounded by
$O$. It can be correct on clips where both branches err (69 to 227 clips per
cell here), so $U$ can in principle be negative. It is positive in every cell we test
(Table~\ref{tab:main}).
Throughout, $O$ is the two-way
construction over the audio branch and the visual ensemble.

We compare three fusion arms. \textbf{Concat} trains one head on the
concatenated frozen embeddings. \textbf{Oracle-route} is the infeasible
$O$ arm, included as the ceiling. \textbf{Learned-router} trains a
selector on the concatenated audio and visual embeddings together with
per-branch confidence statistics ($1539$ dimensions) to predict audio
errors and defer those clips to the visual head; it is an instance of
learning to defer~\cite{mozannar2020defer} and of selective
prediction~\cite{geifman2017selective}, with abstention replaced by
deferral to a second branch.

\section{Results}
\label{sec:results}

\begin{table*}[t]
\centering
\caption{Oracle headroom against realized fusion gain. All rows use the
same configuration (audio branch plus visual ensemble, concatenation head); $H$, $\Delta$, $U$, and $\rho=\Delta/H$ (with 95\% actor-bootstrap CI) are as defined in
Sec.~\ref{sec:defs}. Headroom grows $3.4\times$ from contaminated to
clean audio; the converted fraction stays at or below $0.55$ ($0.39$, $0.55$,
$0.12$). All quantities are three-seed means (seeds $\{0,1,2\}$); $H$,
$\Delta$, or $\rho$ recomputed from the rounded cells can differ by one in
the last digit.}
\label{tab:main}
\small
\setlength{\tabcolsep}{6pt}
\begin{tabular}{lcccccccc}
\toprule
Corpus / audio & audio & video & concat & $O$ & $H$ & $\Delta$ & $U$ & $\rho$ [95\% CI] \\
\midrule
CREMA-D / contaminated & 0.898 & 0.652 & 0.919 & 0.953 & $+$0.055 & $+$0.021 & 0.033 & 0.39 [0.29, 0.48] \\
CREMA-D / clean        & 0.686 & 0.652 & 0.787 & 0.871 & $+$0.186 & $+$0.102 & 0.084 & 0.55 [0.47, 0.62] \\
EAV / contaminated     & 0.648 & 0.460 & 0.665 & 0.785 & $+$0.137 & $+$0.017 & 0.120 & 0.12 [$-$0.02, 0.25] \\
\bottomrule
\end{tabular}
\end{table*}

\subsection{CREMA-D under the contaminated audio encoder}
Concatenation improves on the audio branch from $0.898$ to $0.919$, significant for every visual branch
(per-actor paired Wilcoxon, $n=91$, all $p<10^{-5}$).
The oracle ceiling, however, does not move with the visual branch.
Swapping EmoNet for MAE-DFER, for prenormalized MAE-DFER, or for their
ensemble raises standalone visual accuracy from $0.574$ to $0.652$ but
leaves the oracle near $0.95$ and headroom pinned at approximately
$+0.05$ for every branch (seed 0 for the three single encoders).
Paired two-one-sided-tests on per-actor fusion accuracy ($n=91$) confirm
all six pairs of the four visual encoders are statistically equivalent at a
$\pm0.01$ margin~\cite{lakens2017tost}.

\subsection{CREMA-D under the clean audio encoder}
Replacing emotion2vec+ with the self-supervised non-plus encoder drops
audio accuracy by about $0.21$, from $0.898$ to $0.686$, while the
visual control is unchanged.
This drop bounds the contamination inflation from above but also
includes the supervised-versus-self-supervised capability gap, which we
cannot net out because the clean encoder cannot be evaluated on EAV
(Sec.~\ref{sec:setup}). We therefore frame the effect as encoder
replacement, with contamination as one component of the drop.
With clean audio the realized fusion gain
rises from $+0.021$ to $+0.102$ ($4.8\times$, CI $3.7$--$6.7$) and the
oracle headroom from $+0.055$ to $+0.186$ ($3.4\times$).
The converted fraction $\rho$ rises only from $0.39$
to $0.55$ (paired difference $+0.16$, CI $+0.06$ to $+0.26$), so a
$3.4\times$ larger headroom buys a $1.4\times$ larger fraction, and nearly
half the enlarged headroom goes unconverted. Most of the $3.4\times$ reflects the audio branch's larger error mass.
If the branches erred independently, headroom would equal
$(1-\WAR(a))\,\WAR(v)$; the observed $H$ is $83$--$91\%$ of that
prediction in all three cells, and visual accuracy on audio-error clips
($0.54$--$0.59$ on CREMA-D) is below its overall $0.652$. The headroom
thus carries no complementarity beyond what the branch accuracies imply.
The familiar reading of Table~\ref{tab:main}'s first row, that the
visual branch is nearly worthless for fusion, is therefore partly an
artifact of an inflated audio branch, a form of modality
dominance~\cite{peng2022balanced}.

\begin{figure}[t]
\centering
\includegraphics[width=\columnwidth]{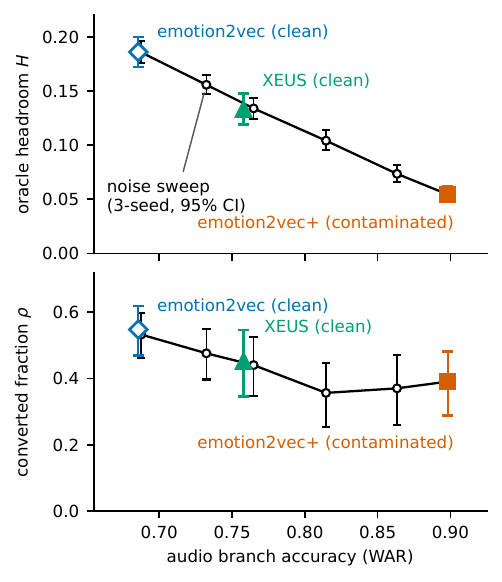}
\caption{Oracle headroom $H$ (top) and converted fraction $\rho$ (bottom)
against audio-branch accuracy on CREMA-D. The line degrades the contaminated
encoder with isotropic Gaussian noise on its audio embeddings (three-seed means, 95\% actor
bootstrap CIs); markers are the contaminated, clean, and XEUS audio
encoders. All three encoders fall on the degradation curve, so headroom
follows audio-branch accuracy rather than encoder identity, while $\rho$
stays at or below $0.55$.}
\label{fig:sweep}
\end{figure}

To separate provenance from branch strength, we hold the contaminated
encoder fixed and add isotropic zero-mean Gaussian noise to its frozen
audio embeddings for all clips, with a fresh draw per seed and fold and
standardization refitted afterwards. Five noise levels, chosen to bring
audio accuracy from $0.90$ down to $0.69$, plus the noise-free baseline
are each retrained with the Table~\ref{tab:main} protocol. Headroom rises steadily from $+0.055$ to $+0.186$ as audio
accuracy falls (Fig.~\ref{fig:sweep}), and at matched audio accuracy the
degraded contaminated encoder reproduces the clean encoder's headroom.
Because isotropic noise makes audio errors close to random, the sweep
alone would sit near the independence prediction by construction; the
real encoders matter because they land on the same curve.
XEUS~\cite{chen2024xeus}, an architecturally distinct self-supervised
encoder (E-Branch\-former, HuBERT objective) whose million-hour pretraining
includes no CREMA-D by any documented pathway, falls on the same curve
($H=+0.133$ at $0.758$~\WAR). Headroom therefore follows audio-branch
accuracy whichever encoder produces it; contamination matters because it
inflates that accuracy. Along the sweep, $\rho$ is non-monotone, dipping
to $0.36$ before rising to $0.53$.

The strongest realized fusion we found is a learned combiner over the
frozen penultimate representations of separately trained audio and visual
heads. Under clean audio it reaches about $0.81$ against
concatenation's $0.787$, significant per actor ($p<0.001$) and in all six
of six seeds, and robust to normalization and depth
controls~\cite{companion2026}.
Even this arm converts only about two thirds of the headroom and leaves
about $0.06$ absolute accuracy on the table, more than the entire headroom available in
the contaminated regime.

\subsection{EAV replication}
EAV is harder and has larger headroom ($+0.137$, Table~\ref{tab:main}).
Concatenation's gain over audio is not significant (per-actor Wilcoxon $p=0.068$, $r=+0.328$, $n=42$).
EAV breaks the link between headroom size and conversion across
corpora. Its headroom is $2.5\times$ CREMA-D's contaminated headroom, yet
its converted fraction, $0.12$ (CI $-0.02$ to $0.25$), is
indistinguishable from zero. A likelier factor than the headroom itself is EAV's weak
visual branch ($0.460$ against $0.652$), below the visual accuracy the one
converting mechanism requires (Sec.~\ref{sec:discussion}); headroom alone
does not reveal this.

\subsection{Selection does not realize the ceiling}
Under contaminated CREMA-D audio the learned router reaches $0.883$, below both concatenation
($0.919$; per-actor Wilcoxon $p<0.001$, $r=-0.71$) and audio alone
($0.898$). Against audio it gains 82 clips and loses 192, a net of $-110$.
Under clean audio the router reaches $0.693$, far below concatenation
at $0.787$.
On EAV it reaches $0.632$, below
concatenation ($p=0.042$, $r=-0.36$) and a net of $-69$ clips against
audio.
Among twelve single- and two-stage fusion mechanisms
swept in our companion paper~\cite{companion2026}, the learned combiner above is the only
family that beats concatenation, and only under clean audio; the two capacity-matched
single-stage alternatives tie or lose in every cell, so it is the
combiner's two-stage structure, not fusion expressivity in general, that
is load-bearing.

\section{Discussion and Limitations}
\label{sec:discussion}

The binding constraint is how the disagreement region is exploited, not
whether it can be detected. The learned router predicts audio errors only moderately
(AUROC $0.773$ on CREMA-D, $0.621$ on EAV; not directly comparable, as
the shipped routers predict different targets). The failure is not
one of gate quality. Across every router input variant in the companion
ablation~\cite{companion2026}, richer inputs improve discrimination (AUROC
$+2.4$ to $+15.5$ points) and calibration (ECE $2.4$- to $11.7$-fold
lower) in all three cells, yet no variant beats concatenation. This is
consistent with the hard-deferral policy, rather than the gate's inputs,
being the bottleneck.
We hypothesize that unconditional concatenation does better because it
declines the identification problem, letting the head learn a fixed
mixture rather than a per-clip decision.

Several limitations qualify the result. All encoders are frozen
extractors with a trained head and both corpora are acted or elicited
studio recordings, so no spontaneous in-the-wild condition is tested. We evaluate
three cells because no clean bimodal corpus is
available to us. The candidate third corpora, IEMOCAP and MELD, are
EmoBox members and additionally lie in the non-plus encoder's
pretraining set, so neither is clean under either audio encoder; and both have multi-party
video without reliable per-utterance speaker-face association, which we
verified directly for MELD.
The combiner wins on CREMA-D but not EAV for a measurable reason rather
than a corpus quirk. A second controlled sweep~\cite{companion2026} fixes
audio degradation where the combiner already wins and degrades only the visual branch. The advantage needs visual accuracy above roughly
$0.54$--$0.60\,\WAR$ and vanishes below it ($p=0.15$ at $0.46\,\WAR$).
On EAV, even the best visual configuration reaches only $0.48\,\WAR$. The combiner needs both
a degraded audio branch and an adequate visual one, and EAV supplies only
the first.

\section{Conclusion}
\label{sec:conclusion}

Oracle complementarity is not realizable complementarity. Replacing a
fine-tuned audio encoder with a self-supervised one triples the oracle
headroom on CREMA-D ($+0.055$ to $+0.186$), and a controlled degradation
shows the increase follows audio-branch accuracy. The converted fraction
rises only from $0.39$ to $0.55$, and on EAV, whose headroom is larger
than the contaminated one, it is indistinguishable from zero. Explicit selection is beaten by plain
concatenation in every corpus and regime tested. Oracle ceilings should
be reported as diagnostics of branch disagreement rather than as fusion targets, and complementarity claims should report $\rho$ alongside $O$, compare $H$
with the independence prediction $(1-\WAR(a))\,\WAR(v)$, and
check whether the evaluation corpus appears anywhere in each encoder's
training lineage.

\section*{Acknowledgments}
The author used Claude (Anthropic) for language editing and revision
throughout the paper.
All AI-assisted content was reviewed and verified by the author, who takes
full responsibility for the paper.

\section*{Compliance with Ethical Standards}
This work is a secondary analysis of the publicly available CREMA-D and
EAV corpora, both distributed for research use by their originating
institutions. No new human-subject data were collected and no additional
ethical approval was required for the analyses reported here.

\bibliographystyle{IEEEbib}

\end{document}